\pdfoutput=1
\documentclass[sigconf]{acmart}

\usepackage{booktabs} 
\usepackage{url}

\usepackage{color}
\usepackage{enumitem}
\usepackage{xcolor}
\usepackage{listings}
\usepackage{graphicx} 
\usepackage{multirow}
\usepackage{siunitx}
\usepackage{rotating}
\usepackage{eqparbox}
\usepackage{wrapfig}
\usepackage{mathtools}
\usepackage{xspace}
\usepackage{enumitem}
\usepackage{hyperref}
\usepackage{lstlinebgrd}
\usepackage{amssymb}
\usepackage{amsmath}
\usepackage{hhline}
\usepackage{wrapfig}

\usepackage{lstautogobble}
\usepackage[framemethod=tikz]{mdframed}
\usetikzlibrary{shadows}
\newmdenv[tikzsetting= {fill=white!20},roundcorner=10pt, shadow=true]{myshadowbox}
\usepackage{graphics}
\usepackage{colortbl} 
\usepackage{multirow} 
\usepackage{balance}
\usepackage{picture}
\usepackage{soul}
\usepackage{array}
\usepackage{makecell}
\usepackage{censor} 

\usepackage{times}
\usepackage{wasysym}

\usepackage{xspace}
\definecolor{ScarletRed}{rgb}{0.80,0.00,0.00}

\definecolor{ForestGreen}{HTML}{00F9DE}

\usepackage{verbatim}
\usepackage{algorithm}
\usepackage{algorithmicx}
\usepackage{algpseudocode}
\usepackage[export]{adjustbox}
\renewcommand{\footnotesize}{\scriptsize}
\usepackage{pgfplots}
\usepackage{pgfplotstable}
\pgfplotsset{compat=newest}
\usepackage{caption}

\usepackage{array}

\newcommand{\calcfactor}[1]{%
  \dimexpr#1\textwidth-2\tabcolsep-1.5\arrayrulewidth\relax
}
\newcolumntype{P}[1]{p{\calcfactor{#1}}}

\setcopyright{rightsretained}

\usepackage{verbatim}
\usepackage{algorithm}
\usepackage{algorithmicx}
\usepackage{algpseudocode}

\definecolor{MyDarkBlue}{rgb}{0,0.08,0.45} 
\acmDOI{XX.YY/ZZ}

\acmISBN{ZZ-YY-24-ZZ/QQ/A}

\acmConference[MSR'18]{Sweden}{May 2018}{} 
\acmYear{2018}
\copyrightyear{2018}

\acmPrice{15.00}

\acmSubmissionID{123-A12-B3}

\begin{document}

\copyrightyear{2018} 
\acmYear{2018} 
\setcopyright{acmcopyright}
\acmConference[MSR '18]{MSR '18: 15th International Conference on Mining Software Repositories }{May 28--29, 2018}{Gothenburg, Sweden}
\acmBooktitle{MSR '18: MSR '18: 15th International Conference on Mining Software Repositories , May 28--29, 2018, Gothenburg, Sweden}
\acmPrice{15.00}
\acmDOI{10.1145/3196398.3196442}
\acmISBN{978-1-4503-5716-6/18/05}

\title{Data-Driven Search-based Software Engineering}

\author{Vivek Nair, Amritanshu Agrawal, Jianfeng Chen, Wei Fu, George Mathew, Tim Menzies, Leandro Minku$^*$, Markus Wagner$^+$, Zhe Yu}
\affiliation{%
  \institution{North Carolina State University, USA; $^*$University of Leicester,UK; $^+$The University of Adelaide, Australia}
}
\email{}

\begin{abstract}
This paper introduces Data-Driven Search-based Software Engineering (DSE), which combines insights from Mining Software Repositories (MSR) and Search-based Software Engineering (SBSE). While MSR formulates software engineering problems as data mining problems, SBSE reformulate Software Engineering (SE) problems as optimization problems and use meta-heuristic algorithms to solve them. Both MSR and SBSE share the common goal of providing insights to improve software engineering. The algorithms used in these two areas also have intrinsic relationships.  
We, therefore, argue that combining these two fields
is useful for situations (a)~which require learning from a large data
source or (b)~when optimizers need to know the lay of the land to find better solutions, faster.

This paper aims to answer the following three questions: (1) What are the various topics addressed by DSE?, (2) What types of data are used by the researchers in this area?, and (3) What research approaches do researchers use? The paper briefly sets out to act as a practical guide to develop new DSE techniques and also to serve as a teaching resource.

This paper also presents a  resource (tiny.cc/data-se) for exploring DSE.  The resource contains 89 artifacts which are related to DSE,
divided into 13 groups such as requirements engineering, software product lines, software processes.
All the materials in this repository
have been used in recent software engineering papers; i.e., for all this material, there exist baseline results against which researchers can comparatively assess their new ideas.
\end{abstract}

%
%



\maketitle

\section{Introduction}

The MSR  community has benefited enormously from widely shared datasets.  Such datasets document the canonical problems in a field.
They serve to cluster together like-minded researchers while allowing them to generate reproducible results. Also, such datasets can be used by newcomers to learn the state-of-the-art methods in this field. Further, they enable the mainstay of good science:
reputable, repeatable,  improvable, and even refutable results.
At the time of this writing,   nine of the 50 most cited papers in the last 5 years at IEEE Transactions of Software Engineering {all draw their case studies
from a small number of readily available defect prediction datasets.}

As a field evolves, so too should its canonical problems. One reason for the emergence of   MSR  in 2004 was the existence of a new generation of widely-available data mining algorithms that could scale to interestingly large problems. In this paper, we pause to reflect on what other kinds of algorithms are widely available and could be applied to MSR problems.  Another take on the emergence of MSR is that underlying datasets became much more readily available with the emergence of large scale, online SE platforms (SourceForge in 1999, Launchpad in 2004, Google Code Project
in 2006, StackOverflow and GitHub in 2008).
 Specifically, we focus on
{\em search-based software engineering} (SBSE) algorithms. Whereas:
\begin{itemize}[leftmargin=*]
\item
An MSR researcher might deploy a
data miner to learn a model from  data   that predicts for (say) a single target class;
\item
An SBSE researcher might deploy a multi-objective optimizer
to find what {solutions} score best
on multiple target variables.
\end{itemize}

{In other words, both MSR and SBSE share the common goal of providing insights to improve software engineering. However, MSR formulates software engineering problems as data mining problems, whereas SBSE formulates software engineering problems as (frequently multi-objective) optimization problems.}

{The similar goals of these two areas, as well as the intrinsic relationship between data mining and optimization algorithms, has been recently inspiring an} increase in methods that combine MSR and SBSE.
A recent {NII Shonan Meeting}  on {\em Data-driven Search-based Software
Engineering} (goo.gl/f8D3EC)   was well attended by over two dozen senior members of the MSR community.
The workshop concluded that (1) mining software repositories could be improved using {tools from the SBSE community}; and that (2) search-based methods can be enhanced using tools from the MSR community. 
For example:
\begin{itemize}[leftmargin=*]
\item
MSR data mining algorithms can be used to summarize the data, after which SBSE can leap to better solutions, faster~\cite{krall2015gale}.
\item
SBSE {algorithms} can be used to select intelligently 
settings for MSR data mining algorithms (e.g.
such as how many trees should be included in a random
forest~\cite{fu2016tuning}).
\end{itemize}
The workshop also concluded that this community needs more shared resources to help more researchers and educators explore MSR+SBSE.
Accordingly, this paper describes  tiny.cc/data-se,
a collection of artifacts for exploring
DSE (see Figure~1). 
All the materials in this repository
have been used in recent SE papers; i.e., for all this material, there exist baseline results against which researchers can use to assess their new ideas comparatively.

It has taken several years to build this resource. Before its
existence, we explored DSE
prototypes on small toy tasks that proved uninteresting to reviewers from SE venues. 
Recently we have much more success in terms of novel
research results
(and publications at SE forums) after 
expanding that collection
to include models
of, e.g., software product lines, requirements, and
agile software projects. 
As such we have found it to be a handy resource
which we now offer to the MSR community.

The rest of this paper discusses SBSE and its connection to MSR. 
We offer a ``cheats' guide'' to SBSE with just enough information to help researchers and educators use the artifacts in the resource. More specifically, the contributions of the paper are:
\begin{itemize}[leftmargin=*]
    \item To show that optimization (SBSE) and learning (MSR) goes hand in hand (Section~\ref{sec:why}),
    \item Provide resources to seed various research (Section~\ref{sec:scenarios}),
    \item Provide teaching resources, which can be used to create DSE courses (Section~\ref{sec:scenarios}),
    \item Based on our experience, various strategies which can be used to make these DSE techniques more efficient (Section~\ref{sec:guide}), and
    \item List of open research problems to seed further research in this area (Section~\ref{sec:open}). 
\end{itemize}

\begin{figure}[t]
 \setlength{\fboxsep}{0.5pt}
 
\fbox{ \includegraphics[width=3.2in]{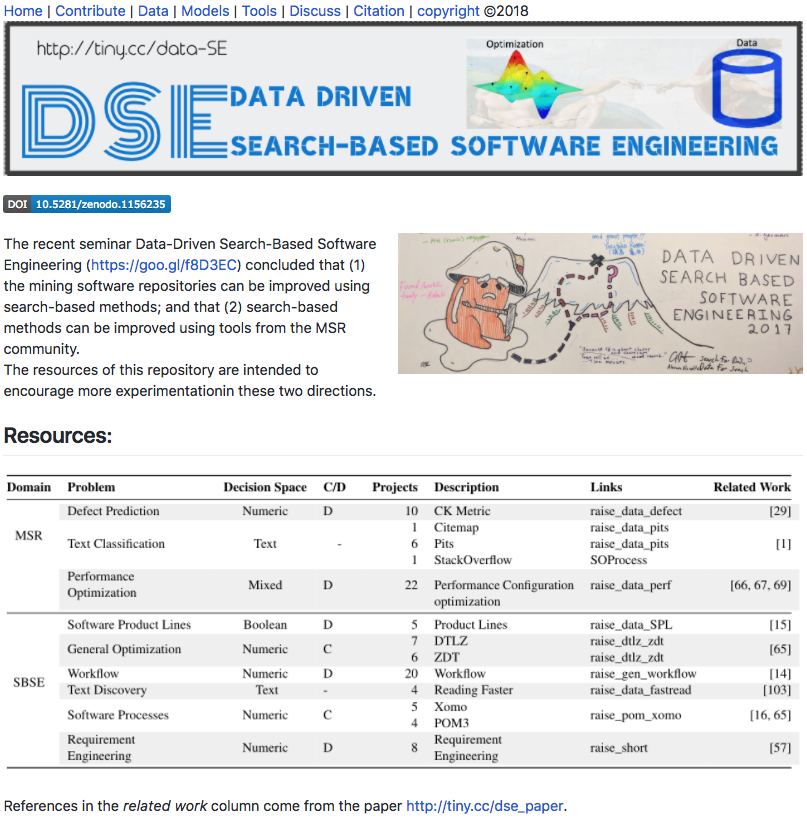}}
\caption{{\small http://tiny.cc/data-se}\label{fig:one}}
\end{figure}

\begin{figure*}
        \centering
        \small
        \colorbox{gray!10}{
            \begin{tabular}{@{}p{3cm}p{6cm}p{8.35cm}@{}}
                \multicolumn{3}{c}{\cellcolor{gray}\textcolor{white}{\textbf{Problems in Search-based Software Engineering}}} \\ 
                \multicolumn{3}{c}{
                        \begin{minipage}[b]{0.33\linewidth}
                            \begin{flushleft}
                                \includegraphics[height=4.5cm]{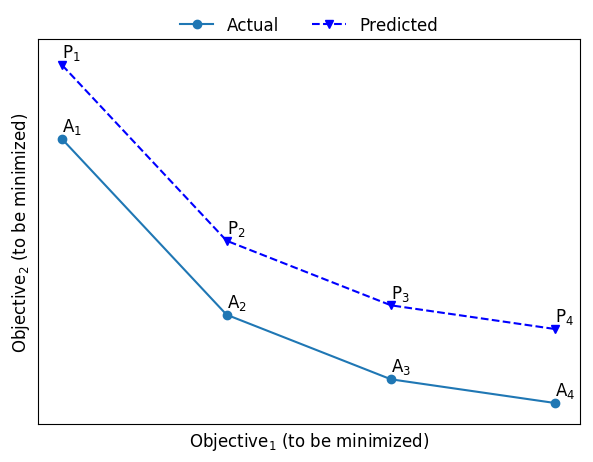}
                                \vspace{8pt}
                            \end{flushleft}
                        \end{minipage}\hspace{0.1cm}
                        \begin{minipage}[b]{0.63\linewidth}
                                \textbf{Problem: } SBSE converts a SE problem into an optimization problem, where the goal is to find the maxima or minima of objective functions {$y_i=f_i(x)$, $1 \leq i \leq m$, where  $f_i$ are the objective / evaluation functions, $m$ is the number of objectives,} x is called the \textit{independent variable}, and $y_i$ are the \textit{dependent variables}.\\
                                \textbf{Global Maximum/Minimum: } For single objective ($m=1$) problems, algorithms aim at finding a single solution able to optimize the objective function, i.e., a global maximum/minimum. \\
                                \textbf{Pareto Front: } For multi-objective ($m>1$) problems, there is no single `best' solution, but a number of `best' solutions. The best solutions are {non-dominated solutions} found using a \textit{dominance} relation.\\
                                \textbf{Dominance: }The domination criterion  can be defined as: ``A solution $x_1$ is said to dominate\\
                                another solution $x_2$, if $x_1$ is no worse than $x_2$ in all objectives
                                and $x_1$ is strictly better than $x_2$ in at least one objective.'' A solution is called non-dominated if no other solution dominates it.\\
                                \textbf{Actual {Pareto Front} (PF): } A list of best solutions of a space is called Actual PF ($a\in A$). As this can be unknowable in practice or prohibitively expensive to generate, it is common to take from the union of all optimization outcomes all non-dominated solutions and use it as the (approximated) PF.\\ 
                                \textbf{Predicted PF: } The solutions found by {an optimization} algorithm are called the Predicted PF ($p\in P$).
                        \end{minipage}
                } \\ 
                \multicolumn{3}{c}{\cellcolor{gray}\textcolor{white}{\textbf{Components of Meta-heuristic Algorithms (search-based optimization algorithms such as NSGA-II~\cite{deb2000fast}, SPEA2~\cite{zitzler2001spea2}, MOEA/D~\cite{zhang2007moea}, AGE~\cite{wagner2015age})}}} \\
                \multicolumn{3}{c}{
                \begin{minipage}[b]{\linewidth}
                \vspace{0.1cm}
                \begin{wrapfigure}{r}{0.4\textwidth}
                \includegraphics[height=4.8cm]{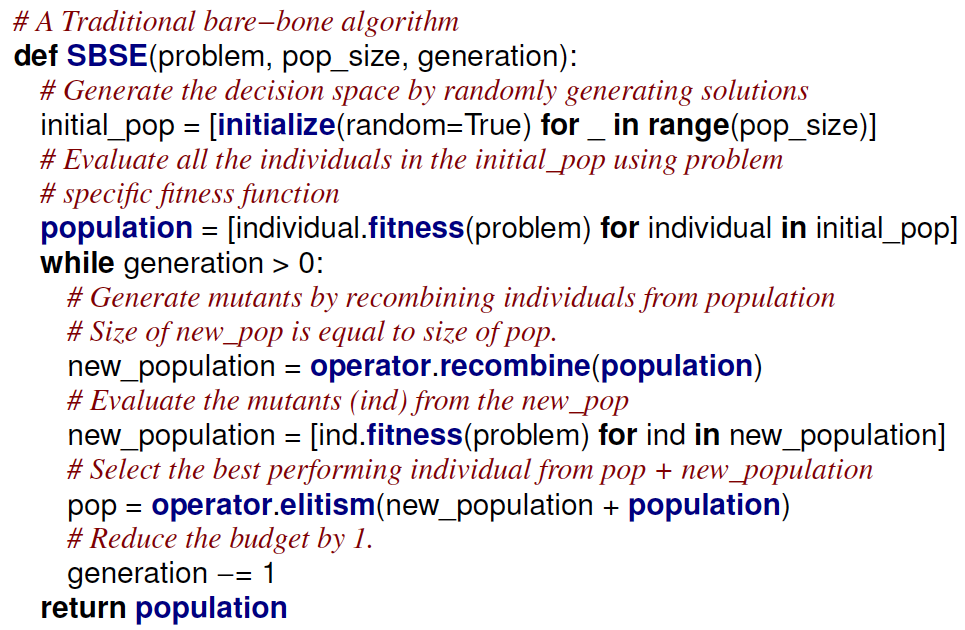}
                \end{wrapfigure}
                \textbf{Data Collection or Model building: } SBSE process can either rely on a model, which represent a software process~\cite{boehm1995cost} or can be directly applied to any software engineering problem including problems which require evaluating a solution by running a specific benchmark~\cite{krall2015gale}. 
                
                \textbf{Representation: } {This defines how solutions $x$ are represented internally by the algorithm.
                Examples of representations are Boolean or numerical vectors, but more complicated representations are also possible. The space of all values that can be represented is the} \textit{Decision Space}.
                
                \textbf{Population: } A set of solutions maintained by the algorithm using the representation.
                
                \textbf{Initialization: } The process of search {typically} starts by creating random solutions (valid or invalid)~\cite{saber2017seeding, chen2017beyond, chen2017sampling, henard2015combining}.
                
                \textbf{Fitness Function: } A fitness function maps the solution (which is represented using numerics) to a numeric scale (also called as \textit{Objective Space}) which is used to distinguish between good and not so good solutions. This measure is a domain-specific measure (single objective) or measures (multi-objective) which is useful for the practitioners.
                Simply put fitness function is a transformation function which converts a point in the decision space to the objective space. 
                
                \textbf{Operators: } {These are operators that (1) generate new solutions based on one (e.g., mutation operator) or more (e.g., crossover or recombination operators) existing solutions, and (2) operators that select solutions to pass to the aforementioned operators or to survive for the next iteration of the algorithm. The operators (2) typically apply some pressure towards selecting better solutions either deterministically (e.g., elitism operator) or stochastically.} 
                \textit{Elitism operator} simulates the `survival of the fittest' strategy, i.e., eliminates not so good solutions thereby preserving the good solutions in the population. 
                
                \textbf{Generations: } {A meta-heuristic algorithm iteratively improves the population (set of solutions) iteratively. Each step of this process, which includes generation of new solutions using recombination of the existing population and selecting solutions using the elitism operator, is called a generation.  Over successive generations, the population `evolves' toward an optimal solution.}
                \end{minipage}\hspace{0.1cm}
                }
                 \\
                \multicolumn{3}{c}{\cellcolor{gray}\textcolor{white}{\textbf{Performance Measures (Refer to \cite{wang2016practical,chand2015manyemo} for more detail)}}} \\ 
                \multicolumn{3}{c}{
                        \begin{minipage}[b]{0.33\linewidth}
                                \includegraphics[height=2.8cm]{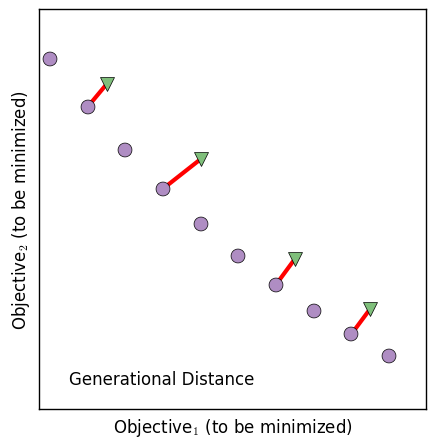}
                                \includegraphics[height=2.8cm]{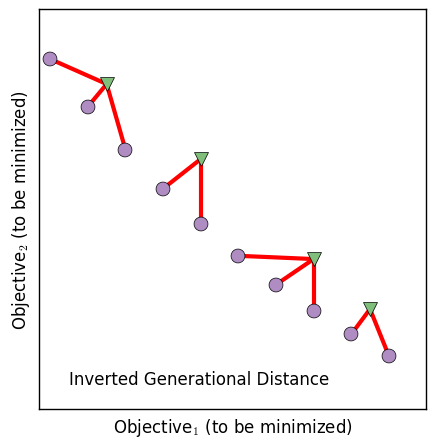}\\
                                \includegraphics[height=2.8cm]{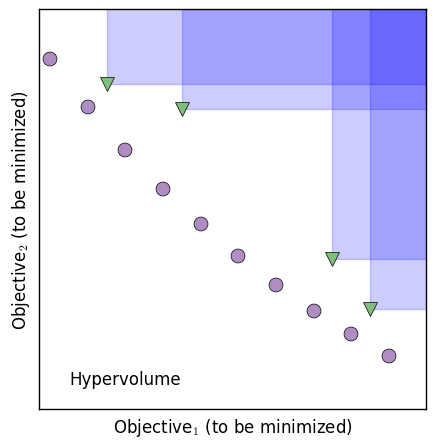}
                                \includegraphics[height=2.8cm]{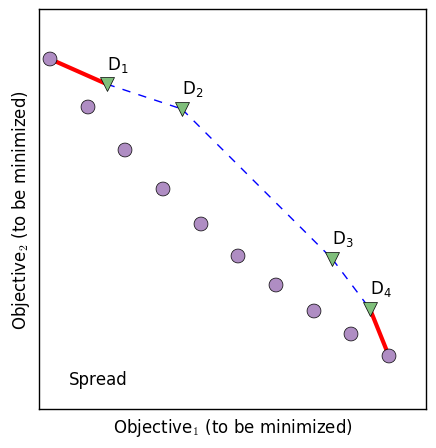}
                                \vspace{0.5cm}
                        \end{minipage}
                        \begin{minipage}[b]{0.66\linewidth}
                            \vspace{0.1cm}
                            For single objective problems, measures such as \textit{absolute residual} or \textit{rank-difference} can be very useful and cannot be used for multi-objective problems. The following are the measures used for such problems.\\
                            \textbf{Generational Distance: } Generational distance is the measure of convergence---how close is the predicted Pareto front is to the actual Pareto front. It is defined to measure (using Euclidean distance) how far are the solutions that exist in $P$ from the nearest solutions in $A$. In an ideal case, the GD is 0, which means the predicted PF is a subset of the actual PF. Note that it ignores how well the solutions are spread out.
                            
                            \textbf{Spread: } Spread is a measure of diversity---how well the solutions in P are spread. An ideal case is when the solutions in P is spread evenly across the Predicted Pareto Front. 
                            
                            \textbf{Inverted Generational Distance: } Inverted Generational distance measures both convergence as well as the diversity of the solutions---measures the shortest distance from each solution in the Actual PF to the closest solution in Predicted PF. Like Generational distance, the distance is measured in Euclidean space. In an ideal case, IGD is 0, which means the predicted PF is same as the actual PF.
                            
                            \textbf{Hypervolume: } Hypervolume measures both convergence as well as the diversity of the solutions---hypervolume is the union of the cuboids w.r.t. to a reference point. Note that the hypervolume implicitly defines an arbitrary aim of optimization. Also, it is not efficiently computable when the number of dimensions is large, however, approximations exist.
                            
                            \textbf{Approximation: } Additive/multiplicative Approximation is an alternative measure which can be computed in linear time (w.r.t. to the number of objectives). It is the multi-objective extension of the concept of approximation encountered in theoretical computer science.

                        \end{minipage}    
                }
            \end{tabular}
        }
        \caption{{\small A {brief} tutorial on Search-based Software Engineering.}}
        \label{fig:sbse_crash}
\end{figure*}

\begin{figure*}[t]
    \centering
    \small
    \begin{tabular}{@{}p{3cm}p{6cm}p{8.35cm}@{}}
                
                \toprule
                \textbf{Features} & \hspace{2cm}\textbf{MSR} & \hspace{3cm}\textbf{SBSE} \\ \midrule
                \rowcolor[HTML]{EFEFEF} \textbf{Primary inference} & Induction, summarization, visualization & Optimization \\
                \textbf{Inference speed} & Very fast, scalable & Becoming faster, more scalable\\
                \rowcolor[HTML]{EFEFEF}\textbf{Data} often collected &
                 Once,  then processed.
                &
                On-demand
                from  some model---which means an SBSE analysis can 
                generate new samples of the data in very  specific regions.
                \\ 
                \textbf{Conclusions} often 
                generated via
                &
                A single  execution of a data miner or an ensemble, 
                perhaps after some data pre-processing. 
                &
                An evolutionary process  where, many times, execution $i+1$ is guided
                by the results of execution $i$.
                \\ 
                 
                \rowcolor[HTML]{EFEFEF}\textbf{Canonical   tools} are usually
                &
                Data mining algorithms like  the decision
                tree learners of WEKA~\cite{hall2009weka} or the mathematical modeling tools of R~\cite{rmanual18} or Scikit-learn~\cite{scikit-learn}.
                &
                {Search-based} optimization algorithms which may be either home-grown scripts or parts of large toolkits such as jMETAL\cite{refs2jmetalDEE}, Opt4j~\cite{opt4jpaper}, DEAP~\cite{DEAP_JMLR2012}.
                \\ 
                \textbf{Canonical   problems}
                &
                Defect prediction~\cite{lessmann2008benchmarking}   or Stackoverflow text mining~\cite{fu2017easy}. 
                &
                Minimizing a test suite~\cite{fraser2007redundancy}, configuring a software system~\cite{nair2017faster} or extracting a valid products from a product line description~\cite{sayyad13b}.
                \\ 
                
                \rowcolor[HTML]{EFEFEF} \textbf{Results}   assessed by
                &
                A small number of standard measures
                including   recall, precision, or false alarm rates (for discrete classes)
                or magnitude of relative error (for continuous classes)~\cite{Shepperd2012}.
                &
                A wide-variety of domain-specific objectives.
                Whatever the specific objectives, a small number meta-measures are used in many research papers such as the Hypervolume or Spread. 
    \end{tabular}
    \caption{{\small Differences between MSR and Search-based Software Engineering.}}
    \label{fig:diff}
\end{figure*}

 \begin{figure}[b]
\small

\begin{tabular}{p{.95\linewidth}}\hline 
\rowcolor{gray!10}
Some goals relate to aspects of defect prediction:
\begin{enumerate}[leftmargin=0.4cm]
 
\item
Mission-critical systems are risk-averse and may accept very high false alarm rates, just as long as they catch any life-threatening possibility. That is, such projects
do not care about effort- they want to {\em maximize recall} regardless of any impact
that might have on the false alarm rate.
\item  Suppose a new hire wants to impress their manager. That the new hire might want to ensure that no result presented to  management contains  true negative;
i.e., they wish to {\em maximize precision}.
\item
Some communities do not care about low precision, just as long as a small fraction the data is returned. Hayes, Dekhytar, \& Sundaram call this fraction
{\em selectivity} and offer an extensive discussion of the merits of this measure~\cite{hayes06}.
\end{enumerate}
\\ \hline
\rowcolor{gray!10}
Beyond defect prediction are other goals that combine defect prediction with other economic
factors:
\begin{enumerate}[leftmargin=0.4cm]
\setcounter{enumi}{3}
\item
Arisholm~\&~Briand~\cite{arisholm06},  Ostrand \& Weyeuker~\cite{ostrand04} and Rahman et al.~\cite{rahman12}
say that a defect predictor should maximize {\em reward}; i.e., find the fewest lines of code
that contain the most bugs.
\item In other work, Lumpe et al. are concerned about
 {\em amateur  bug fixes}

~\cite{me11f}.
Such amateur fixes are highly correlated to errors and, hence, to
avoid such incorrect bug fixes; we have to optimize
for finding the most number of bugs in regions that {\em the most programmers have worked with before}.
\item In {\em better-faster-cheaper} setting, one seeks  project changes that lead
to fewer defects and faster development times using fewer resources~\cite{elrawas10,me07f,me09a,me09f}.
\item
Sayyad~\cite{sayyad13a,sayyad13b} explored models of software product
lines whose value propositions are defined by five objectives.
\end{enumerate}
\\ \hline
\rowcolor{gray!10}
All the above measures relate to the tendency of a predictor to find something. Another measure could be {\em variability} of the predictor.
\begin{enumerate}[leftmargin=0.4cm]
\setcounter{enumi}{7}
\item
In their study on reproducibility of SE results,
 Anda, Sjoberg and Mockus advocate using the coefficient of variation ($CV=\frac{stddev}{mean}$).
Using this measure, they defined {\em reproducibility} as $\frac{1}{CV}$~\cite{mockus09}.
\end{enumerate}\\\hline
\end{tabular}
\caption{{\small Different users value different things.}}\label{fig:goals}
\end{figure}

\section{What? (Definitions)}

For this paper, we say that MSR covers the technologies
explored by many of the authors at the annual Mining Software Repositories
conference.

As to search-based software engineering, that term was coined by Jones and Harman~\cite{harman2001search} in 2001.
Over the years SBSE has been applied to various fields of software engineering for example, requirements~\cite{ZhangHL13, chen2017beyond}, automatic program repair~\cite{le2012genprog}, Software Product Lines~\cite{chen2017sampling, sayyad13a, guo2017smtibea}, Performance configuration optimization~\cite{nair2017faster,nair2017using, guo2017data, oh2017finding, nair2018finding} to name of few. SBSE has been applied to other fields and has their own surveys such as design~\cite{raiha2010survey}, model-driven engineering~\cite{boussaid2017survey}, genetic improvement of programs~\cite{petke2017genetic}, refactoring~\cite{mariani2017systematic}, Testing~\cite{silva2017systematic, khari2017extensive} as well as more general surveys~\cite{clarke2003reformulating, harman2007current}. 

 Figure~\ref{fig:sbse_crash} provides a (very) short tutorial on SBSE and  Figure~\ref{fig:diff} characterizes some of the differences between MSR 
and SBSE.
 
As to defining {\em DSE},
we say it is some system that solves an SE problem as follows:
\begin{itemize}[leftmargin=*]
\item
It inserts  a data miner into an optimizer; or
\item
It uses an optimizer  to improve a data  miner.
\end{itemize}

\section{  Why? (Synergies of MSR + SBSE)} \label{sec:why}

Consider the following. The MSR
community knows how to deal with large datasets. The SBSE community knows how to take
clues from different regions of data, then combine them to generate better solutions.
If the two ideas were combined, then  MSR could be  {\em sampling} technology to quickly
find the regions where SBSE  can learn {\em  optimizations}. If so:
\begin{quote}
\centering
{\bf {\em   MSR  methods can ``supercharge''    SBSE.}}
\end{quote}
For a concrete example of this ``supercharging'', consider active learning
with Gaussian Process Models (GPM) used for multi-objective optimization
by the machine learning community~\cite{zuluaga2016varepsilon}.  Active learners assume that evaluating one candidate
is very expensive. For example, in software engineering, ``evaluating'' a test suite
might mean re-compiling the entire system then re-running all tests. When the evaluation is so slow,
an active learner reflects on the examples seen so far to find the next most informative
example to evaluate. One way to do this is to use GPMs to
find which parts of a model have maximum variance in their predictions 
(since sampling
in such high-variance regions serves to most constrain the model).  

The problem with GPMs is that they do not scale beyond a dozen variables (or features)~\cite{wang2016bayesian}. CART, on the other hand, is a data mining algorithm that scales efficiently
to hundreds of variables. So Nair et al. recently explored active learning for multi-objective optimization by replacing GPM with one CART tree per objective~\cite{nair2018finding}.
 The resulting system was applied to a wide range of software configuration problems
 found in \url{tiny.cc/data-se}.
 Compared to GPMs, the resulting system ran orders of magnitude faster, found solutions as good or better, and scaled to much larger problems~\cite{nair2018finding}. 

Not only is   MSR useful for   SBSE, but so too:
\begin{quote}
\centering
{\bf {\em SBSE methods can    ``supercharge''    MSR.}}
\end{quote}
The standard example here is parameter tuning. Most data mining algorithms come with tunable parameters that have to be set via expert judgment. Much recent work shows that for
MSR problems such as defect prediction and text mining, SBSE can automatically find settings that are far superior to the default settings. For example:
\begin{itemize}[leftmargin=*]
 \item
   When performing defect prediction, various groups report that SBSE methods can find new settings that dramatically improve the performance of the learned model~\cite{fu2016tuning,tantithamthavorn2016automated, Tantithamthavorn2018}.

\item When using SMOTE to rebalance data classes, SBSE found that for 
distance calculations using 
  \[d(x,y)=\left(\sum_i(x_i-y_i)^n\right)^{1/n}\] the Euclidean distance of $n=2$ usually works far worse than another distance measure using $n=3$~\cite{agrawal2017better}.
 
\end{itemize} 
Note that all the above-used case study material is from tiny.cc/data-se.

 Another, subtler, benefit of combining MSR+SBSE relates to the exploration of competing
 objectives.
 We note that software engineering tasks rarely involve a single goal. For example, when a software engineer
is testing a software, he/she may be interested in finding the highest possible number of software defects at the same time as minimizing the time required for testing. Similarly, when a software
engineer is planning the development of a software,
he/she may be interested in reducing the number of
defects, the effort required to develop the software and the cost of the software. The existence of {\em
multiple goals} necessarily implies that such problems should be solved via a multi-objective optimizer.

There are many such goals. For example, 
let $\{A,B,C,D\}$ denote the
true negatives,
false negatives,
false positives, and
true positives
(respectively) found by a software defect detector.
Also, let $L_A L_b, L_c, L_d$ be the lines of code
seen in the parts of the system that fall
into $A, B, C, D$. Given these definitions then

{\small\[
\begin{array}{r@{~}l}
\mathit{pd}=\mathit{recall}=&D / (B+D)\\
\mathit{pf}=&C / (A+C)\\
\mathit{prec}=\mathit{precision}=&D / (D+C)\\
\mathit{acc}=\mathit{accuracy}=&(A+D) / (A+B+C+D)\\
\mathit{support}=&(C+D) / (A+B+C+D)\\
\textit{effort}=&(L_c+L_d) / (L_a + L_b + L_c + L_d)\\
\mathit{reward}=&\mathit{pd} / \textit{effort}
\end{array}
\]}.

The critical point here is that, in terms of evaluation criteria,  the above are just the tip of the iceberg.
Figure~\ref{fig:goals} lists several other criteria that have appeared recently in the literature. Note that this list is hardly complete-- SE has many sub-problems and many of those
problems deserve their specialized evaluation criteria.

SBSE is one way to build inference systems that are specialized to specialized evaluation
criteria. SBSE systems accept as input some function that assesses examples on multiple
criteria (and that function is used to guide the inference of the SBSE tool).
Several recent results illustrate the value of using a wider range of evaluation criteria
to assess our models:
\begin{itemize}[leftmargin=*]
\item
Sarro et al.~\cite{sarro2016multi}  used SBSE tools that assessed software effort estimation tools
not only by their predictive accuracy but also by the confidence of those predictions.
These multi-objective methods out-performed the prior state-of-the-art in effort estimation. 
{Prior work also used multi-objective methods to boost the predictive performance of ensembles for
software effort estimation~\cite{minku2013}.}
\item
Agrawal et al.~\cite{agrawalwrong} used SBSE tools to tune text mining tools for StackOverflow.
Standard text mining tools can suffer from ``order effects'' in which changing the order
of the training data leads to large-scale changes in the learned clusters.
To fix this, Agrawal et al. tuned the background priors
of their Latent Dirichlet allocation (LDA) algorithm to maximize the stability
of the learned clusters.  Classifiers based on these stabilized clusters performed better than those based on the clusters learned by standard LDA.
\end{itemize}

\begin{table*}[t]
\centering
\small
\caption{{\small Different problems and associated strategies explored in this paper. The characteristic of the decision space (C/D) represents whether there are continuous or discrete in nature. The column Links represent the URL from where the problems can be download (prefix http://tiny.cc/).}}
\label{tbl:only1}
\begin{tabular}{@{}cp{3cm}cp{0.7cm}rp{3cm}lr@{}}
\toprule
\textbf{Domain} & \textbf{Problem} & \textbf{Decision Space} & \textbf{C/D} & \textbf{Projects} & \textbf{Description} & \textbf{Links} & \textbf{Related Work} \\ \midrule
 & \cellcolor[HTML]{EFEFEF}Defect Prediction & \cellcolor[HTML]{EFEFEF}Numeric & \cellcolor[HTML]{EFEFEF}D & \cellcolor[HTML]{EFEFEF}10 & \cellcolor[HTML]{EFEFEF}CK Metric & \cellcolor[HTML]{EFEFEF}\href{http://tiny.cc/raise_data_defect}{raise\_data\_defect} & \cellcolor[HTML]{EFEFEF}\cite{fu2016tuning} \\
 &  &  & \multicolumn{1}{c}{} & 1 &Citemap & \href{http://tiny.cc/raise_data_pits}{raise\_data\_pits} &  \\
 &  &  & \multicolumn{1}{c}{} & 6 & Pits & \href{http://tiny.cc/raise_data_pits}{raise\_data\_pits} &  \\
\multirow{-4}{*}{MSR} & \multirow{-3}{*}{Text Classification} & \multirow{-3}{*}{Text} & \multicolumn{1}{c}{\multirow{-3}{*}{-}} & 1 & StackOverflow & \href{http://tiny.cc/SOProcess}{SOProcess} & \multirow{-3}{*}{\cite{agrawalwrong}} \\
 & \cellcolor[HTML]{EFEFEF}\begin{tabular}[c]{@{}l@{}}Performance\\Optimization\end{tabular} & \cellcolor[HTML]{EFEFEF}Mixed & \cellcolor[HTML]{EFEFEF}D & \cellcolor[HTML]{EFEFEF}22 & \cellcolor[HTML]{EFEFEF}Performance Configuration optimization & \cellcolor[HTML]{EFEFEF}\href{http://tiny.cc/raise_data_perf}{raise\_data\_perf} & \cellcolor[HTML]{EFEFEF}\cite{nair2017faster, nair2017using, nair2018finding} \\
\midrule
 & Software Product Lines & Boolean & D & 5 & Product Lines & \href{http://tiny.cc/raise_data_SPL}{raise\_data\_SPL} & \cite{chen2017sampling} \\
 & \cellcolor[HTML]{EFEFEF} & \cellcolor[HTML]{EFEFEF} & \cellcolor[HTML]{EFEFEF} & \cellcolor[HTML]{EFEFEF}7 & \cellcolor[HTML]{EFEFEF}DTLZ & \cellcolor[HTML]{EFEFEF}\href{http://tiny.cc/raise_dtlz_zdt}{raise\_dtlz\_zdt} & \cellcolor[HTML]{EFEFEF} \\
 & \multirow{-2}{*}{\cellcolor[HTML]{EFEFEF}General Optimization} & \multirow{-2}{*}{\cellcolor[HTML]{EFEFEF}Numeric} & \multirow{-2}{*}{\cellcolor[HTML]{EFEFEF}C} & \cellcolor[HTML]{EFEFEF}6 & \cellcolor[HTML]{EFEFEF}ZDT & \cellcolor[HTML]{EFEFEF}\href{http://tiny.cc/raise_dtlz_zdt}{raise\_dtlz\_zdt} & \multirow{-2}{*}{\cellcolor[HTML]{EFEFEF}\cite{nair2016accidental}} \\
 & Workflow & Numeric & D & 20 & Workflow & \href{http://tiny.cc/raise_gen_workflow}{raise\_gen\_workflow} & \cite{chen2017riot} \\
 & \cellcolor[HTML]{EFEFEF}Text Discovery &\cellcolor[HTML]{EFEFEF} Text & \cellcolor[HTML]{EFEFEF}- & \cellcolor[HTML]{EFEFEF}4 & \cellcolor[HTML]{EFEFEF}Reading Faster & \cellcolor[HTML]{EFEFEF}\href{http://tiny.cc/raise_data_fastread}{raise\_data\_fastread} & \cellcolor[HTML]{EFEFEF}\cite{yu2016read} \\
 &  &  &  & 5 & Xomo &  &  \\
 & \multirow{-2}{*}{Software Processes} & \multirow{-2}{*}{Numeric} & \multirow{-2}{*}{C} & 4 & POM3 & \multirow{-2}{*}{\href{http://tiny.cc/raise_pom_xomo}{raise\_pom\_xomo}} & \multirow{-2}{*}{\cite{nair2016accidental, chen2017beyond}} \\
\multirow{-9}{*}{SBSE} & \cellcolor[HTML]{EFEFEF}{\begin{tabular}[c]{@{}l@{}}Requirement\\ Engineering\end{tabular}} & \cellcolor[HTML]{EFEFEF}Numeric & \cellcolor[HTML]{EFEFEF}D & \cellcolor[HTML]{EFEFEF}8 &
\cellcolor[HTML]{EFEFEF}{\begin{tabular}[c]{@{}l@{}}Requirement\\ Engineering\end{tabular}} & \cellcolor[HTML]{EFEFEF}\href{http://tiny.cc/raise_short}{raise\_short} & \cellcolor[HTML]{EFEFEF}\cite{mathew2017shorter} \\ \bottomrule
\end{tabular}
\end{table*}

\section{How? (Resources for Exploring DSE)}\label{sec:scenarios}
In this section, we assume that the reader
has been motivated by the above material
to start exploring DSE.

Accordingly, in this section, we describe sample tasks that might be used for that exploration.
Note that:
\begin{itemize}[leftmargin=*]
\item
Table~\ref{tbl:only1} summarizes the material presented in this section. 
\item All these examples were used in recent SE
papers; i.e., for all this material, there exists baseline results against which researchers can use to assess their own new ideas comparatively.
\item
Support material for that exploration (data, models, scripts),
is available from \url{tiny.cc/data-se}.

\end{itemize}
Note that we do not assume that the following list of materials
covers the spectrum of problems that could be solved by combining MSR with SBSE. In fact, one of the goals of this paper is to encourage researchers to extend this material by posting their materials as pull requests to \url{tiny.cc/data-se}.

    \subsection{Software Product Lines}\label{spl}
    \textbf{\underline{Problem Domain}: } SBSE
    
    \noindent\textbf{\underline{Problem}:} With fast-paced development cycle, traditional code-reuse techniques have been infeasible. Now, software companies are moving a software product line model to reduce cost and increase reliability. Companies concentrate on building software out of core components, and quickly use these components with specializations for certain customers. This allows the companies to have a fast turn around time. In a more concrete sense, a software product line (SPL) is a collection of related software products, which share some core functionality~\cite{harman2014search}. From a product line, many products can be generated. 
    
\begin{figure}[!t]
    \small
    \includegraphics[width=\linewidth]{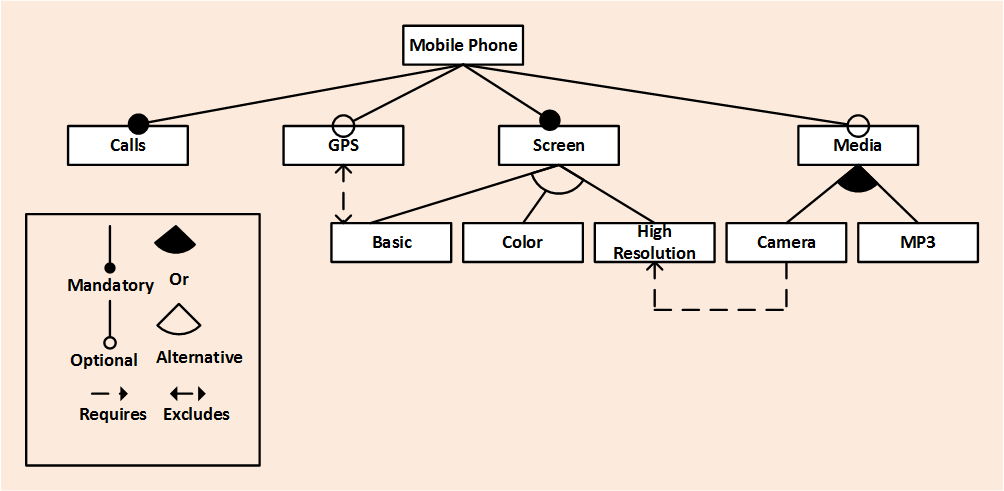}
    \caption{{\small Feature model for mobile phone product line. To form a mobile phone, ``Calls'' and ``Screen'' are the mandatory features(shown as \textit{solid $\bullet$}), while the ``GPS'' and ``Media'' features are optimal(shown as \textit{hollow $\circ$}). The ``Screen'' feature can be ``Basic``, ``Color'' or ``High resolution'' (the \textit{alternative} relationship). The ``Media'' feature contains ``camera'', ``MP3'', or both (the \textit{Or} relationship).}}
    \label{fig:mobile}
\end{figure}

Figure \ref{fig:mobile} shows a feature model for a mobile phone
product line. All features are organized as a tree. The relationship
between two features might be ``mandatory'', ``optional'',
``alternative'', or ``or''. Also, there exist some cross-tree constraints, which means the preferred features are not in the same sub-tree. These cross-tree constraints complicate the process of exploring feature models.\footnote{Without cross-tree constraints, one can generate products in linear time using a top-down traversal of the feature model.} The products are \textit{represented} as a binary string, where the length of the string is equal to the number of features. A valid product is one which satisfies all the relationship defined in the feature model.
Researchers who explore these kinds of models~\cite{sayyad13a, sayyad13b, harman2014search, henard2015combining}
define a ``good'' product (\textit{fitness function}) as the one that satisfies five objectives:
(1) find the valid products (products not violating any cross-tree constraint or tree structure), (2) with more features, (3) less known defects, (4) less total cost, and (5) most features used in prior applications.

\noindent\textbf{\underline{Challenges}: } Finding a valid product in real-world software product lines can be very difficult due to the sheer scale of these product lines. Some software product line models comprise up to tens of thousands
of features,  with 100,000s of constraints. These constraints make it difficult to generate valid product through random assignments. In some cases, chances of finding valid solutions through random assignment are $0.04\%$. Most of the meta-heuristic algorithms often fail to find valid solutions or take a long time to find one. Given the large and constrained search space ($2^N$, where N is the number of features) using a meta-heuristic algorithm can be infeasible.

\noindent\textbf{\underline{Strategy}:} Since exploring all the possible solutions is expensive and often infeasible, the SWAY, or ``Sampling WAY'', clusters the individual products based on their features. Please note, clustering does not require evaluations---to find the fitness of each product. SWAY uses a domain-specific distance function to cluster the points. A domain-specific distance function was required because (1) clusters should have similar products---similar fitness values, and (2) the decision space is a boolean space. This is in line with the observation of Zhang et al.~\cite{zhang2013software}, who reports that MSR practitioners understand the data using domain knowledge. Once the products are clustered, a product is selected (at random) from each cluster. Based on the fitness values of the `representative product,' the not so promising clusters are eliminated. This step is similar to the \textit{elitism} operator of meta-heuristic algorithms. This process continues recursively till a certain budget is reached. Please refer to \cite{nair2016accidental, chen2017beyond, chen2017sampling} for more details. The reproduction package of SWAY and associated material can be found in \url{http://tiny.cc/raise_spl}.

    \subsection{Performance Configuration Optimization}
\textbf{\underline{Problem Domain}: } MSR

\noindent\textbf{\underline{Problem}: } Modern software systems come with a lot of knobs or configuration options, which can be tweaked to modify the functional or non-functional (e.g., throughput or runtime) requirements. Finding the best or optimal configuration to run a particular workload is essential since there is a significant difference between the best and the worst configurations. Many researchers report that modern software systems come with a daunting number of configuration options~\cite{xu2015hey}. The size of the configuration space increases exponentially with the number of configuration options. The long runtimes or cost required to run benchmarks make this problem more challenging.

\noindent\textbf{\underline{Challenges}: } Prior work in this area used a machine learning method to accurately model the
configuration space. 
The model is built sequentially, where new configurations are sampled randomly, and the quality or accuracy of the model is measured using a holdout set. The size of the holdout set in some cases could be up to 20\% of the configuration space~\cite{nair2017using} and need to be evaluated (i.e., measured) before even the machine learning model is entirely built. This strategy makes these methods not suitable in a practical setting since the generated holdout set can be (very) expensive. On the other hand, there are software systems for which an accurate model cannot be built. 

\noindent\textbf{\underline{Strategy}: } The problem of finding the (near) optimal configuration is expensive and often infeasible using the current techniques. A useful strategy could be to build a machine learning model which can differentiate between the good and not so good solutions. Flash, a Sequential Model-based Optimization (SMBO), is a useful
strategy to find extremes of an unknown objective. Flash is
efficient because of its ability to incorporate prior belief as
already measured solutions (or configurations), to help direct
further sampling. Here, the prior represents the already
known areas of the search (or performance optimization) problem. The prior can be used to estimate the rest of the
points (or unevaluated configurations). Once one (or many) points are evaluated based on the prior,
the posterior can be defined. The posterior captures the updated belief in
the objective function. This step is performed by using a
machine learning model, also called surrogate model. 
The concept of Flash can be simply stated as:
\begin{itemize}[leftmargin=*]
\item Given what one knows about the problem,
\item what can be done next?
\end{itemize}
The ``given what one knows about the problem'' part is
achieved by using a machine learning model whereas ``what can be done next'' is performed by an acquisition function.
Such acquisition function automatically adjusts the \textit{exploration}
(``should we sample in uncertain parts of the search
space'') and \textit{exploitation} (``should we stick to what is already
known'') behavior of the method. Please refer to  \cite{nair2018finding} and  \cite{nair2017using,nair2017faster, jamshidi2016uncertainty} to similar strategies. The reproduction package is available in \url{http://tiny.cc/flashrepo/}.

    \subsection{Requirements Models}
\textbf{\underline{Problem Domain}: } SBSE
    
\begin{figure}[!t] 
  ~~~\includegraphics[width=2.7in]{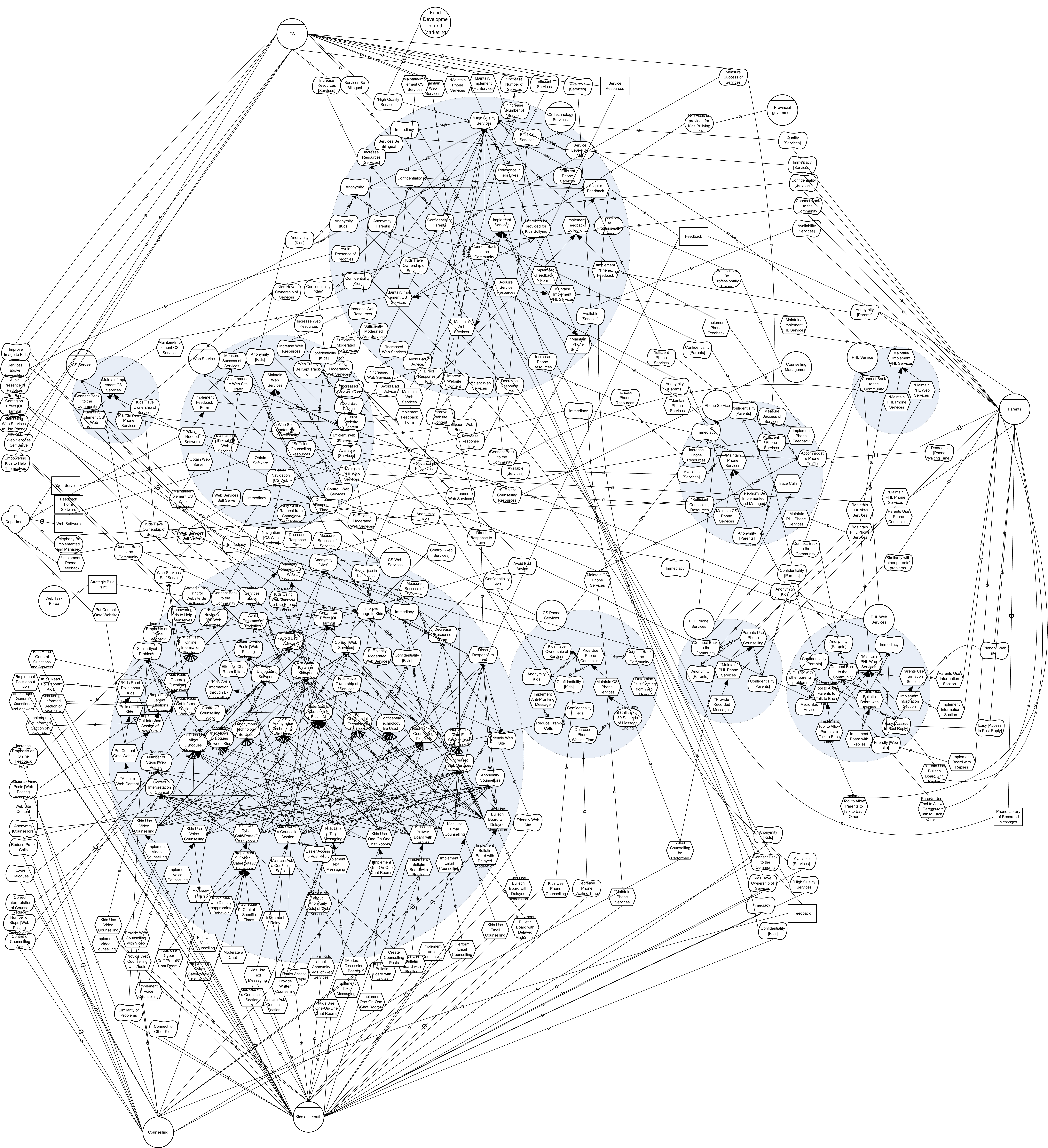} 
    \caption{{\small Options for services in a CS department (i* format). It shows how complicated it is to reason about services.}}
    \label{fig:csServices}
\end{figure}

    \noindent\textbf{\underline{Problem}:} The process of building and analyzing complex requirements engineering models can help stakeholders better understand the ramifications of their decisions~\cite{Lamsweerde2001,amyot10}. But models can sometimes overwhelm stakeholders. For example, consider a committee reviewing a goal model (see fig. \ref{fig:csServices}) that describes the information needs of a computer science department~\cite{Horkoff2016}. Although the model is entangled, on manual and careful examination, it can be observed that much of the model depends on a few ``key'' decisions such that once their values are assigned, it becomes very simple to reason over the remaining decisions. It is beneficial to look for these ``keys'' in requirements models since, if they exist, one can achieve ``shorter'' reasoning about RE models, where ``shorter'' is measured as follows:
    \begin{itemize}[leftmargin=*]
     \item{Large models can be processed in a short time.}
     \item{Runtimes for automatic reasoning about RE models are shorter so stakeholders can get faster feedback on their models.}
     \item{The time required for manual reasoning about models is shorter since stakeholders need only debate a small percent of the issues (just the key decisions).}
    \end{itemize}

Such models are represented using the \textit{i*} framework \cite{yu97a} which include the key concepts of NFR~\cite{mylopoulos92.nfr} framework, including soft goals, AND/OR decompositions and contribution links along with goals, resources, and tasks. 

\noindent\textbf{\underline{Challenges}: } Committees have trouble with manually reasoning about all the conflicting relationships in models like fig. \ref{fig:csServices} due to its sheer size and numerous interactions (this model has 351 node, 510 edges and over 500 conflicting relationships). Further, automatic methods for reasoning about these models are hard to scale up: as discussed below, reasoning about inference over these models is an NP-hard task. 

\noindent\textbf{\underline{Strategy}:} To overcome this problem, a technique called SHORT was proposed, which runs in four phases:
\begin{itemize}[leftmargin=*]
    \item{\textbf{SH}: 'S'ample 'H'euristically the possible labelings of the model.}
    \item{\textbf{O}: 'O'ptimize the label assignments to cover more goals or reduce the sum of the cost of the decisions in the model.}
    \item{\textbf{R}: R: 'R'ank all decisions according to how well they performed during the optimization process.}
    \item{\textbf{T}: T: 'T'est how much conclusions are determined by the decisions that occur very early in that ranking.}
\end{itemize}

The above technique was used on eight large real-world Requirements Engineering models. It was shown that only under 25\% of their decisions are ``keys'' and 6 of them had less than 12\% decisions as keys. The process of identifying keys was also fast as it could run in near linear time with the largest of models running in less than 18 seconds. Please refer to \cite{mathew2017shorter} for more details and the reproduction package can be found at \url{http://tiny.cc/raise_short}.
\vspace{-0.1cm}
\subsection{Faster Literature Reviews}
\textbf{\underline{Problem Domain}: } SBSE

\noindent\textbf{\underline{Problem}: }
 Broad and complete literature reviews.
 
 Data sets and reproduction packages for all the following are available at \url{https://doi.org/10.5281/zenodo.837298} (for Challenge~1) and 
\url{https://doi.org/10.5281/zenodo.1147678} (for Challenges 2,3,4). Please refer to \cite{yu2016finding, YuM17} for more details
   
\vspace{1.0ex}
\noindent\textbf{Challenge 1: }
Literature reviews can be extremely labor intensive and often require months (if not year) to complete. Due to a large number of papers available, the relevant papers are hard to find. A graduate student needs to review thousands of papers before finding the few dozen relevant ones.
Therefore the challenge is: how to maximize relevant information (when there is a lot of noise) while minimizing the search cost to find relevant information?

\noindent\textbf{\underline{Strategy}: }
  Reading all literature is unfeasible. To selectively read the most informative literature, active learners are applied---which incrementally learns from the human feedback and suggests on which paper to review next~\cite{YuM17a}.
   
\vspace{1.0ex}
\noindent\textbf{Challenge 2: }
A wrong choice of initial papers can increase the review effort by up to 300\% than the median effort (repeat for 30 runs)---requires three times more effort to find relevant papers~\cite{yumen17a}. 

\noindent\textbf{\underline{Strategy}: }
Reduce variances by selecting a good initial set of papers with domain knowledge from the researcher. It is found that by ranking the papers with some keywords (provided as domain knowledge) and reviewing in such order, the effort can be reduced with negligible variances~\cite{YuM17}.
   
\vspace{1.0ex}
\noindent\textbf{Challenge 3: }
When to stop?. If too early then many relevant papers will be missed
else much time will be wasted on irrelevant papers. 

\noindent\textbf{\underline{Strategy}: }
Use a semi-supervised machine learning algorithm (Logistic Regression) to learn from the search process (till now) to predict how much more relevant paper will be found~\cite{YuM17}. 

\vspace{1.0ex}
\noindent\textbf{Challenge 4: }
Research shows that it is reasonable to assume the precision and recall of a human reviewer are around 70\%. When such human errors occur, how to correct the errors so that the active learner is not misled?

\noindent\textbf{\underline{Strategy}: }
Concentrate the effort on correctly classifying the paper which creates the most controversy. Using this intuition, periodically few of the already evaluated papers, whose labels the active learner disagree most on, are re-evaluated~\cite{YuM17}.

\vspace{-0.1cm}
    \subsection{Text classification}
\textbf{\underline{Problem Domain}: } MSR

\noindent\textbf{\underline{Problem}: } Stack Overflow is a popular Q\&A website, where users posts the questions and the community collectively answers these questions. However, as the community evolves, there is a chance that duplicate questions can appear---which results in a wasted effort of the community. There is a need to remove the duplicate question or consolidate related questions. The problem focuses on discovering the relationship between any two questions posted on Stack Overflow and classifies them into duplicates, direct link, indirect link, and isolated~\cite{fu2017easy, xu2016predicting}. One way to solve this problem is to build a predictive model to predict the similarity between two questions. 

\noindent\textbf{Challenge: } The state-of-the-art method for this problem used Deep Learning, which was expensive to train~\cite{xu2016predicting}. For example, Xu et al. spent 14 hours on training a deep learning model. Such long training time is not appropriate for the field of software analytics since software analytics requires the methods to have a fast turnaround time~\cite{zhang2013software}.

\noindent\textbf{\underline{Strategy}: }To reduce the training time as well as promote simplicity, hyper-parameter optimization of simple learners, like SVM (Support Vector Machines), is a way to go. Specifically, a {meta-heuristic algorithm called} Differential Evolution, which has been an effective tuning algorithm (used in SBSE)~\cite{fu2016tuning},
was used to explore the parameter space of SVM. After the search process,
 SVM with best-found parameters is used to predict classes of Stack Overflow questions. This method can get similar or better results
than the deep learning method while reducing the training time by up to 80 times. Please refer to \cite{fu2017easy} for more details and the reproduction package can be found in~\url{http://tiny.cc/raise_data_easy}.

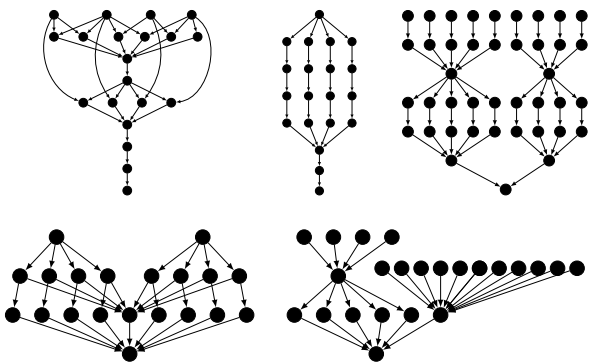
\begin{figure}[!b]
\small
    \centering
    \noindent
    \begin{tabular}{cccccc}
        \resizebox{!}{70pt}{
        \begin{tikzpicture}
[lineDecorate/.style={-latex,line width=0.2mm}, nodeDecorate/.style={shape=circle,inner sep=3pt,draw}]

 \foreach \nodename/\x/\y in {
0/0/0,
1/1/0,
2/2/0,
3/3/0,
4/1.5/-0.5,
5/1.5/-1.0,
6/1.5/-1.5,
7/1.5/-2,
8/-1/2,
9/0.8/2,
10/2.3/2,
11/3.7/2,
12/-1/1.5,
13/0/1.5,
14/1.2/1.5,
15/2.1/1.5,
16/3/1.5,
17/3.9/1.5,
18/1.5/1.0,
19/1.5/0.5
 }
 {
          \node (T\nodename) at (\x,\y*1.5) [nodeDecorate, fill=black] {};
 }
		
\path
\foreach \startnode/\endnode in {0/4,1/4,2/4,3/4,4/5,5/6,6/7,
8/12,8/13,9/12,9/13,9/14,9/15,10/14,10/16,11/15,11/16,11/17,
12/18,13/18,14/18,15/18,16/18,17/18,
18/19,
19/0,19/1,19/2,19/3}
        {
          (T\startnode) edge[lineDecorate] node {} (T\endnode)
        }

(T8) edge[lineDecorate, bend right=50] node{} (T0)
(T10) edge[lineDecorate, bend left=30] node{} (T2)
(T9) edge[lineDecorate, bend right=30] node{} (T1)
(T11) edge[lineDecorate, bend left=70] node{} (T3);
        \end{tikzpicture}}
         
& 
\resizebox{!}{70pt}{
\begin{tikzpicture}
[lineDecorate/.style={-latex,line width=0.2mm}, nodeDecorate/.style={shape=circle,inner sep=3pt,draw}]

 \foreach \nodename/\x/\y in {
0/1.5/0,
1/1.5/-0.5,
2/1.5/-1
 }
 {
          \node (T\nodename) at (\x*0.8,\y*1.5) [nodeDecorate, fill=black] {};
 }

		 \foreach \x in {0,1,2,3}
 {
          \node (A\x) at (\x*0.8, 4.0) [nodeDecorate, fill=black] {};
		  \node (B\x) at (\x*0.8, 3.0) [nodeDecorate, fill=black] {};
		  \node (C\x) at (\x*0.8, 2.0) [nodeDecorate, fill=black] {};
		  \node (D\x) at (\x*0.8, 1.0) [nodeDecorate, fill=black] {};
 }
 
 \node(H) at (1.5*0.8, 5) [nodeDecorate, fill=black]{};
 
\path
\foreach \startnode/\endnode in {0/1,1/2}
        {
          (T\startnode) edge[lineDecorate] node {} (T\endnode)
        }
\foreach \startnode/\endnode in {H/A1,H/A2,H/A3,H/A0}
        {
          (\startnode) edge[lineDecorate] node {} (\endnode)
 }
 
 \foreach \i in {0,1,2,3}
        {
          (A\i) edge[lineDecorate] node {} (B\i)
		  (B\i) edge[lineDecorate] node {} (C\i)
		  (C\i) edge[lineDecorate] node {} (D\i)
		  (D\i) edge[lineDecorate] node {} (T0)
 };

\end{tikzpicture}}

 \hspace{.2in}\resizebox{!}{70pt}{
\begin{tikzpicture}
[lineDecorate/.style={-latex,line width=0.2mm}, nodeDecorate/.style={shape=circle,inner sep=3pt,draw}]

 \foreach \i in {0,1,2,3,4,5,6,7,8}
 {
          \node (A\i) at (\i*0.6,5*0.8) [nodeDecorate, fill=black] {};
		  \node (B\i) at (\i*0.6,4*0.8) [nodeDecorate, fill=black] {};
		  \node (E\i) at (\i*0.6,2*0.8) [nodeDecorate, fill=black] {};
		  \node (F\i) at (\i*0.6,1*0.8) [nodeDecorate, fill=black] {};
 }
 \node(C) at (2*0.6,3*0.8) [nodeDecorate, fill=black]{};
 \node(D) at (6.5*0.6,3*0.8) [nodeDecorate, fill=black]{};
 \node(G) at (2*0.6,0*0.8) [nodeDecorate, fill=black]{};
 \node(H) at (6.5*0.6,0*0.8) [nodeDecorate, fill=black]{};
  \node(I) at (4.5*0.6,-1*0.8) [nodeDecorate, fill=black]{};
\path

 \foreach \i in {0,1,2,3,4,5,6,7,8}
        {
          (A\i) edge[lineDecorate] node {} (B\i)
		  (E\i) edge[lineDecorate] node {} (F\i)
 }
  \foreach \i in {0,1,2,3,4}
        {
          (B\i) edge[lineDecorate] node {} (C)
		  (C) edge[lineDecorate] node {} (E\i)
		  (F\i) edge[lineDecorate] node {} (G)
 }
  \foreach \i in {5,6,7,8}
        {
          (B\i) edge[lineDecorate] node {} (D)
		  (D) edge[lineDecorate] node {} (E\i)
		  (F\i) edge[lineDecorate] node {} (H)
 }

 (G) edge[lineDecorate] node {} (I)
 (H) edge[lineDecorate] node {} (I);
\end{tikzpicture}}

\\
~\\

\resizebox{!}{50pt}{
\begin{tikzpicture}
[lineDecorate/.style={-latex,line width=0.2mm}, nodeDecorate/.style={shape=circle,inner sep=3pt,draw}]

 \foreach \i in {0,1,2,3}
 {
          \node (A\i) at (\i*0.6+0.15,1*0.8) [nodeDecorate, fill=black] {};
		  \node (B\i) at (\i*0.6,0*0.8) [nodeDecorate, fill=black] {};
		  
 }
 
  \foreach \i in {5,6,7,8}
 {
          \node (A\i) at (\i*0.6-0.15,1*0.8) [nodeDecorate, fill=black] {};
		  \node (B\i) at (\i*0.6,0*0.8) [nodeDecorate, fill=black] {};
		  
 }
 
 \node(X) at (1.5*0.6, 2*0.8) [nodeDecorate, fill=black]{};
 \node(Y) at (6.5*0.6, 2*0.8) [nodeDecorate, fill=black]{};
\node(K) at (4*0.6, 0*0.8) [nodeDecorate, fill=black]{};
\node(E) at (4*0.6,-1*0.8) [nodeDecorate, fill=black]{};
\path

 \foreach \i in {0,1,2,3}
        {
          (A\i) edge[lineDecorate] node {} (B\i)
		  (A\i) edge[lineDecorate] node {} (K)
		  (B\i) edge[lineDecorate] node {} (E)
		  (X) edge[lineDecorate] node {} (A\i)
 }
 (K) edge[lineDecorate] node{}(E)
  \foreach \i in {5,6,7,8}
        {
          (A\i) edge[lineDecorate] node {} (B\i)
		  (A\i) edge[lineDecorate] node {} (K)
		  (Y) edge[lineDecorate] node {} (A\i)
		  (B\i) edge[lineDecorate] node {} (E)
 };
 
\end{tikzpicture}}

&

\resizebox{!}{50pt}{
\begin{tikzpicture}
[lineDecorate/.style={-latex,line width=0.2mm}, nodeDecorate/.style={shape=circle,inner sep=3pt,draw}]

 \foreach \i in {0,1,2,3,4}
 {
          \node (A\i) at (\i*0.6,-1*0.8) [nodeDecorate, fill=black] {};
		  
 }
 
  \foreach \i in {0,1,2,3}
 {
          \node (C\i) at (\i*0.6+0.2,1*0.8) [nodeDecorate, fill=black] {};
		  
 }
 
   \foreach \i in {0,1,2,3,4,5,5,6,7,8,9,10}
 {
          \node (F\i) at (\i*0.4+1.8,0.2*0.8) [nodeDecorate, fill=black] {};
		  
 }
 
 \node(B) at (1.5*0.6, 0*0.8) [nodeDecorate, fill=black]{};
 \node(D) at (2.8*0.6, -2*0.8) [nodeDecorate, fill=black]{};
  \node(E) at (5*0.6, -1*0.8) [nodeDecorate, fill=black]{};
\path
(E) edge[lineDecorate] node {} (D)  
 \foreach \i in {0,1,2,3}
        {
          (C\i) edge[lineDecorate] node {} (B)  
 }
  \foreach \i in {0,1,2,3,4}
        {
          (A\i) edge[lineDecorate] node {} (D)
		  (B) edge[lineDecorate] node {} (A\i)
 }
   \foreach \i in {0,1,2,3,4,5,5,6,7,8,9,10}
        {
          (F\i) edge[lineDecorate] node {} (E)
 };
\end{tikzpicture}}

\\
    \end{tabular}
    \caption{
    {\small Some  of  cloud computing  workflows. Clockwise from top left: {\tt Montage, Epigenomics, Inspiral, CyberShake, Sipht}. Each node is one ``task''
    and each edge is a data flow from one task to another.   Number of tasks  can vary from dozens to   thousands).}  
    }
    \label{fig:structure}
\end{figure}
\vspace{-0.1cm}
\subsection{Workflows in Cloud Environment}
\textbf{\underline{Problem Domain}: } SBSE

\noindent\textbf{\underline{Problem}:} Many complex computational tasks, especially in a scientific area, can be divided into several sub-tasks where outputs of some tasks serves as the input to another. A workflow is a tool to
model such kind of computational tasks.

Figure \ref{fig:structure} shows five types of widely studied workflows. Workflows are represented as a directed acyclic graph (DAG). Each vertex of the DAGs represents one sub-task. Connections between vertices represent the result communications between different vertices. One computational task is called ``finished'' only when
all sub-tasks are finished. Also, sub-tasks should follow the constraints per edges.

Grid computing techniques, invented in the mid-1990s, as well as recently lots of pay-as-you-go cloud computing services, e.g., Amazon EC2, provide researchers a feasible way to finish such kind of complicated workflow in a reasonable time.
The workflow configuring problem is to figure out the best deployment configurations onto a cloud environment. The deployment configuration
contains (1) determine which sub-tasks can be deployed into one computation node, i.e., the virtual machine in cloud environment; (2) which sub-task should be executed first if two of them are to be executed in the same node; (3) what hardware configuration (CPU, memory, bandwidth, etc.) should be applied in each computation node. Commercial cloud service provider charges users for computing resource. 
The objective of cloud configuration is to minimize the monetary cost as well as minimize the runtime of computing tasks.

\noindent\textbf{\underline{Challenges}: } 
The two reasons that make workflow configuration on cloud environment challenging are: (1) some sub-tasks of workflows can be as large as hundreds, or thousands; also, sub-tasks are under constraints-- file flow between them. (2) available types of
computing nodes from cloud service providers are large.
For example, Amazon AWS provides more than 50 types of virtual machines; these
virtual machines have different computing ability, as well as unit price (from \$0.02/hr to \$5/hr).
Given these two reasons,
the configuration space, or some possible deployment ways, is large.
Even though modern cloud environment simulator, such as CloudSim, can quickly
access performance of one deployment, evaluate every
possible deployment configuration is impossible.

\noindent\textbf{\underline{Strategy}:} 
Since it is impossible to enumerate all possible configurations, most existed algorithm use either (1) greedy algorithm (2) evolutionary algorithm to figure out best configurations.
Similar to other search-based software engineering problems,
these methods require a large number of model evaluations (simulations in workflow configuration problem).
The RIOT, or randomized instance-order-type, is an algorithm to balance execution time as well as cost
in a short time.
The RIOT first groups sub-tasks, making the DAG simpler, and then
assign each group into one computation node. Within one node,
priorities of sub-tasks are determined by
B-rank~\cite{topcuoglu2002performance}, a greedy algorithm. 
The most tricky part is to determine types of computation nodes so that they can coordinate with each other and reduce the total ideal time (due to file transfer constraints).
RIOT makes full use of two hypothesis: (1) similar configurations should have similar performance and (2) (monotonicity) k times computing resource should lead to (1/k)*c less computation time (where c is constant for one workflow).
With these two hypotheses,
RIOT first randomly creates some deployment configurations and evaluates them, then guess more configurations based on current evaluated configuration. Such kind of guess can avoid a large number of configuration evaluations. Please refer to \cite{chen2017riot} for more details and the reproduction package is available in \url{http://tiny.cc/raise_gen_workflow}.

\vspace{-0.3cm}
\section{Practical Guidelines} \label{sec:guide}

Based on our experiences this section lists practical guidelines for those who are familiar with MSR (but not with SBSE) to be able to work on DSE. 
Though one would expect exceptions or probably simpler techniques, given the current techniques listed in this paper (Section~\ref{sec:scenarios}), we believe that it is essential to provide practical guidance that will be valid in most cases and enable practitioners to use these techniques to solve their problems. 
We recommend that practitioners follow these guidelines and possibly use them as a teaching resource.

\noindent\textbf{Learning: } To learn SBSE, coding up a Simulated Annealer~\cite{van1987simulated} and Differential Evolution (DE)~\cite{storn97} is a good starting point.  These algorithms work well for single-objective problems. For multi-objective problems, one should code up binary domination and indicator domination~\cite{zitzler2001spea2}. Note that the indicator domination is recommended for N>2 objectives and that indicator domination along with DE can find a wide spread of solutions on the PF. 

As to learning more about area, the popular venues are: 
\begin{itemize}[leftmargin=*]
\item \href{http://gecco-2018.sigevo.org/index.html/HomePage}{GECCO}: The Genetic and Evolutionary Computation Conference (which has a special SBSE track);
\item TSE: IEEE Transactions on Evolutionary Computation;
\item SSBSE: the annual symposium on search-based SE;
\item Recent papers   at FSE, ICSE, ASE, etc: \url{goo.gl/Lvb42Z};
\item Google Scholar: see  \url{goo.gl/x44u77}.
\end{itemize}
  \noindent\textbf{Debugging: } It is always recommended to use small and simple {optimization problems} for the purposes of debugging. For example, small synthetic {problems} like ZDT, DTLZ~\cite{deb2005scalable}, and WFG~\cite{huband2006review} are very useful to evaluate a meta-heuristic algorithm. For further details about other test problems, please refer to \cite{huband2006review}.  
  
  That said, it is {\em not} recommended that you try publishing results based on these small {problems}. In our experience, SE reviewers are more interested in results from the resources listed in Table~\ref{tbl:only1}, rather than results from synthetic {problems} like ZDT, DTLZ, and WFG. Instead, it is better to publish results on interesting problems taken from the SE literature, such as those shown in Section \ref{sec:scenarios}.
  
  Another aspect of debugging is to notice the gradual improvement of results in terms of the performance metrics. In most cases, the performance metric for generation $n$ would be worse than generation $n+\delta$, where $\delta$ is a positive integer. In an unstable meta-heuristic algorithm, the performance metrics fluctuate a lot between generations. 
  {It is also common to observe a fitness curve over generations where the fitness initially improves more quickly and then starts to converge. However, a typical undesired result is premature convergence, which occurs when the algorithm converges rapidly too to poor local optimal solutions.}

  \noindent\textbf{Normalization: } When working with multi-objective problems, it is important to normalize the objective space to eliminate scaling issues. For example, Ishibuchi et al.~\cite{ishibuchi2017effect} showed that WFG4-9 test problems~\cite{huband2006review}, the range
of the Pareto front on the $i^{th}$ objective is $[0, 2i]$. This means
that the tenth objective has a ten times wider range than
the first objective. It is difficult for meta-heuristic algorithm without normalization
to find a set of uniformly distributed solutions over the entire Pareto front for such a many-objective test
problem. 

\noindent\textbf{Choosing Algorithm and its parameters: } {Choosing the meta-heuristic algorithm to solve a particular problem can often be challenging. Although the chosen algorithm is a preference of the researcher, it is commonly agreed to use NSGA-II or SPEA2 for problems with less than 3 objectives whereas NSGA-III and MOEA/D is preferred for problems with more than 3 objectives.}

{Choosing the correct parameters of a meta-heuristic algorithm are essential for its performance. There are various rules of thumb, which are proposed by various researchers such as population size is the most significant parameter and that the crossover probability and mutation rate have insignificant effects on the GA performance~\cite{alajmi2014selecting}. However, we recommend using a simple tuner to find the best parameter for the meta-heuristic algorithms. Tuners could be something as simple as Differential Evolution or Grid Search. }

 \noindent\textbf{Experimentation: } We recommend an experimentation technique called (you+two+next+dumb),  defined as 
\begin{itemize}[leftmargin=*]
\item ``you'' is your new method;
    \item ``two''  well-established methods (often NSGA-II and SPEA2),
    \item a ``next''  generation method e.g. MOEA/D~\cite{zhang2007moea}, NSGA-III~\cite{deb2014evolutionary},
    \item  one ``dumb'' baseline method (random search or SWAY).
\end{itemize}

While comparing a new meta-heuristic algorithm or a DSE technique, it is essential to baseline it with the simplest baseline such as a random search. Bergestra et al.~\cite{bergstra2012random} demonstrated that random search which uses same computational budget finds better solutions by effectively searching a larger, less promising configuration space. Another alternative could be to generate numerous random solutions and reject less promising solutions using domain-specific heuristics. Chen et al.~\cite{chen2017sampling}, showed that SWAY---oversampling and subsequent pruning less promising solutions, is competitive with state-of-the-art solutions. Baselines like random search or SWAY can help researchers and industrial practitioners by achieving fast early results while providing `hints' for subsequent experimentation.
Another important aspect is to compare the performance of the new technique with the current state-of-the-art techniques. Well established techniques in the SBSE literature are NSGA-II and SPEA2~\cite{chen2017beyond}. Please note that the state-of-the-art techniques differs among the various sub-domains.

  \noindent\textbf{Reporting results: }
Meta-heuristic algorithms in SBSE are an intelligent modification of a randomized algorithm. Like randomized algorithms, the meta-heuristic algorithms may be strongly affected by chance. Running a randomized
algorithm twice on the same problem usually produces different results. Hence, it is imperative to run the meta-heuristic algorithm multiple times to capture the behavior of an algorithm. Arcuri et al.~\cite{arcuri2011practical} reports that meta-heuristic algorithms should be \textit{run at least 30 times}. Take special care to use \textit{different random seeds} for running each iteration of the algorithms\footnote{We once and accidentally reset the random number seed to "1" in the inner loop of the experimental setup. Hence,  instead of getting 30 repeats with different seeds, we got 30 repeats of the same seed. This lead to two years of wasted research based
on an effect that was just a statistical aberration.}. This makes sure that the randomness is accounted for while reporting the results. 
To analyze the effectiveness of a meta-heuristic algorithm, it is important to study the distribution of its performance metrics. A practitioner might be tempted to use the average (mean) of the performance metrics to compare the effectiveness of different algorithms. However, given the variance of the performance metrics between different runs just looking only at average values can be misleading. For detecting statistical differences and compare central tendencies and overall distributions, \textit{use of non-parametric statistical methods} such as Scott-Knott using bootstrap and cliffs delta for the significance and effect size test~\cite{mittas2013ranking, ghotra2015revisiting}, Friedman~\cite{lessmann2008benchmarking} or Mann-Whitney U-test~\cite{arcuri2011practical}-- please refer to \cite{arcuri2011practical, arcuri2014hitchhiker}.\footnote{
As to statistical methods, our results are often heavily skewed so don't use anything that assumes symmetrical Gaussians-- i.e., no t-tests or ANOVA.}

  \noindent\textbf{Replication Packages: } As a community, we advance by accumulating knowledge built upon observations of various researchers. We believe that replicating an experiment (thereby observations) many times transforms evidence into a trusted result. The goal of any research should be not the running
of individual studies but developing a better understanding
of process and debate about various strength and weakness of the approach.
In the experiments with DSE, there are many uncontrollable sources of variation exist from
one research group to another for the results of any study, no
matter how well run, to be extrapolated to other domains. 

Along with increasing the confidence of a certain observation, it also increases the speed of research. For example, recently in FSE'17, Fu et al.~\cite{fu2017revisiting} described the effects of \textit{arxiv.org} or the open science effect. Fu et al. described how making the paper, and the replication packages publicly available, results in 5 different techniques (each superior to its predecessor). Please see \url{http://tiny.cc/unsup}  for more details on arxiv.org effect. 

Hence, it is essential to make replication packages available. 
In our experience, we have found that replication packages hosted on personal web pages tend to disappear or end up with dead links after a few years.
We have found that storing replication packages and artifacts on Github and registering it with Zenodo (\href{https://zenodo.org}{https://zenodo.org}) is an effective strategy. 
Note that once registered, then every new release (on Github) will be backed up on Zenodo and made available. Also considering posting a link to your package on tiny.cc/data-se, which is a more curated list. 
\vspace{-0.3cm}
\section{Open Research Issues} \label{sec:open}
We hope this article inspires a larger group of researchers to work
on open, and compelling problems
in DSE. There are many such problems, including the few listed below.

  \noindent\textbf{Explanation: } SBSE is often instance-based and provide optimal or (near) optimal solutions. This is not ideal since it provides no insight into the problem. However, using MSR techniques finding
building blocks of good solutions may shed light on the relationship
different parameters that affect the quality of a solution. This is an important shortcoming of SBSE which have not been addressed by this community. Valerdi
notes that, without automated tools, it can take days for human
experts to review just a few dozen examples. In that same
time, an automatic tool can explore thousands to billions of more
solutions. Humans can find it an overwhelming task just to
certify the correctness of conclusions generated from so many
results. Verrappa and Leiter warn that:
``... for industrial problems, these algorithms generate
(many) solutions which make the task of understanding
them and selecting one amongst them difficult
and time-consuming.''~\cite{veerappa2011understanding}. Currently, there has been only a few efforts to comprehend the results of an SBSE technique. Nair et al.~\cite{nair2017flash}
uses a \textit{domination tree} along with a Bayesian-based method to better, and more succinctly,
explain the search space. 

 \noindent\textbf{Human in the loop: } Once we have explanation running, then the next step would be to explore combinations of human and artificial intelligence for
enhanced DSE.
Standard genetic algorithms must evaluate 1000s to 1,000,000s of examples-- which makes it hard for engineers or business users to debug or audit the results of that analysis. On the 
other hand, tools like FLASH (described in the last bullet)
and SWAY (see Section \ref{spl}) only evaluated $O(log(N))$ of the candidates-- which in practice can be just a few dozen examples. This number is small enough to ask humans
to watch the reasoning and, sometimes, catch the SBSE
tool making mistakes as it compares alternatives.  This style of human in the loop reasoning could be used for many tasks such as:
\begin{itemize}[leftmargin=*]
\item Debugging or auditing the reasoning.
\item Enhancing the reasoning; i.e., can human intelligence, applied judiciously,  boost artificial intelligence?
\item Checking for missing attributes: when human experts say two identical examples are different, we can infer that there are extra attributes, not currently being modeled, that are important.
\item Increasing human confidence in the reasoning by tuning, a  complex process, into something
humans can monitor and understand.
\end{itemize}

 \noindent\textbf{Transfer Learning: } The premise of software analytics is that there exists data
from the predictive model can be learned. However in the cases where data is scarce, sometimes it is possible to use data collected from same or other domains
which can be used for building predictive models. There is some recent work exploring the problem of transferring data from one domain to another for data analytics. These research have focused on two methodological variants of transfer learning: (a) dimensionality
transform based techniques~\cite{nam2013transfer, krishna2016too, nam2017heterogeneous,minku2014transfer}
and (b) the similarity based approaches~\cite{kocaguneli2011find, kocaguneli2015transfer, peters2015lace2}.
These techniques can be readily applied to SBSE to reduce further the cost of finding the optimal solutions. For example, while searching for the right configuration for a specific workload ($w_a$), we could reuse measurement from a prior optimization exercise, which uses a different workload ($w_b$), to further prune the search space~\cite{jamshidi2017transfer, jamshidi2017transfer, valov2017transferring}.

\noindent\textbf{Optimizing Optimizers}: 
Many of the methods above can be used to tune approaches. However, many of them experience difficulties once the underlying function is noisy or the algorithm stochastic (and thus the optimizer gets somewhat misleading feedback), once many parameters are to be tuned, or once the approach is to be tuned for large corpora of instances where the objectives vary significantly. Luckily, in recent years, many automated parameter optimization methods have been developed and published as software packages. General purpose approaches include ParamILS~\cite{hutter2007paramils}, SMAC~\cite{hutter2011smac}, GGA~\cite{ansotegui2009gga}), and the iterated f-race procedure called irace~\cite{birattari2002irace}. 
Of course, even such algorithm optimizers can be optimized further. One word of warning: as algorithm tuning is already computationally expensive, the tuning of algorithm tuners is even more so. While Dang et al.~\cite{dang2017iraceconfig} recently used surrogate functions to speed up the optimizer tuning, more research is needed for the optimization of optimizers more widely applicable. Lastly, an alternative to the tuning of algorithms is that of selecting an algorithm from a portfolio or determining an algorithm configuration, when an instance is given. This typically involves the training of machine learning models on performance data of algorithms in combination with instances given as feature data. {In software engineering, this has been recently used as a DSE approach for the Software Project Scheduling Problem \cite{Shen2018,wu2016}.} The field of per-instance configuration has received much attention recently, and we refer the interested reader to a recent updated survey article~\cite{kotthoff2016survey}. 

\vspace{-0.3cm}
\section{Conclusions}
SE problems can be solved both by MSR and SBSE techniques, but both these methods have their shortcomings. This paper has argued these shortcomings can be overcome by merging ideas from both these domains, to give rise to a new field of software engineering called DSE. This sub-area of SE boosts the techniques used in MSR and SBSE by drawing inspiration from the other field. 

This paper proposes concrete strategies which can be used to combine the techniques from MSR and SBSE to solve an SE problem. It also list resources which   researchers  can use to jump-start their research. One of the aims of the paper is to provide resources and material which can be used as teaching or training resources for a new generation of researchers. 

\vspace{-0.3cm}
\section*{Acknowledgements}
This work was inspired by the recent
 NII Shonan Meeting on Data-Driven Search-based Software Engineering (goo.gl/f8D3EC), December 11-14, 2017.
We thank the organizers of that workshop 
(Markus Wagner,  
 Leandro Minku,  
 Ahmed E. Hassan, and 
 John Clark)
for their academic
leadership and inspiration. Dr. Minku\textquotesingle s work has been partly supported by EPSRC Grant No. EP/R006660/1.

\balance
\bibliographystyle{plain}


\end{document}